\documentclass[12pt]{revtex4-1}
\usepackage{amsmath}
\usepackage{amsfonts}
\usepackage{amssymb}
\usepackage{graphicx}
\usepackage{color}     
\usepackage{hyperref} 
\usepackage{longtable}
\usepackage{multirow}

\begin{document}
\title{Quadrupole transitions in the bound rotational-vibrational spectrum 
of the tritium molecular ion}
\author{Horacio Olivares Pil\'on}
\email{horop@nucleares.unam.mx}
\affiliation{Physics Department, University of Connecticut, 2152 Hillside Rd., Storrs, Connecticut 06269-3046, USA}
\affiliation{Departamento de F\'isica, Universidad Aut\'onoma Metropolitana,
    Apartado Postal 55-534, 09340 M\'exico, DF, M\'exico}

\date{\today}
\begin{abstract}
The nonrelativistic energies of the homonuclear  ion T$_2^+$ are
calculated for the ground state using the Lagrange-mesh method as was
done for the isotopomers H$_2^+$ and D$_2^+$ ({\it J. Phys. B:
  At. Mol. Opt. Phys.} {\bf 45} 065101 and {\it J. Phys. B:
  At. Mol. Opt. Phys.} {\bf 46} 245101). Energies and eigenfunctions
are obtained up to four of the lowest  bound vibrational states
($v=0,1,2,3$) which support 62, 61, 60 and 58 bound rotational
states, respectively.  Some quasibound sates are also presented until
$L =$ 68.  From the obtained wave functions, electric quadrupole
transitions per time unit are calculated between those states over the
whole rotational bands. Extensive results are presented  with 6 significant digits. 
The ground state energy of the symmetric systems $^{\infty}$H$_2^+$, $^{\infty}$H$^-$,  Ps$^{-}$, H$^-$, D$^-$, T$^-$, $\mu^+\mu^+e$ and $\mu^+e\,e$ is presented with high accuracy as a function of $\beta=m_3/(2m+m_3)$.
\end{abstract}
\maketitle

\section*{Introduction}

The  molecular ion T$_2^+$ composed of two tritium nuclei (tritons) and one electron is the heaviest isotopomer of H$_2^+$.  This three body  system  is the less studied molecular ion in the family of three-body molecular systems with Coulomb interaction composed of protons, deuterons, tritons  and one electron.  A complete quantitative description has been done is some papers  mainly for the ground state  ($v=0$ and $L=0$).  For this state, a complete list of various geometrical and energetically properties were obtained by Frolov~\cite{Fr99,Fr02}. Polarizabilities and energies were  calculated with high precision  by Yan et al.~\cite{YZL03} not only for the ground state but also for the lowest P state. In 2004 Yan et al.~\cite{YZ04} reported non relativistic energies for those states with vibrational quantum number $v=0$ and  $L$ ranging  from 2 to 12.  However, at present time, the number of  rotational-vibrational bound states supported by the ground state is not known.  

The goal of this paper is to present a systematic description of the spectra for the lowest four vibrational states $v=0,1,2,3$ and all the bound rotational states without considering the Born-Oppenheimer approximation. A few rotational states beyond the dissociation limit are also presented. In order to do that the Lagrange mesh method is applied in the same way that was done for the homonuclear systems H$_2^+$~\cite{OPB12} and D$_2^+$~\cite{OP13}. This method is an approximate variational calculation using a basis of Lagrange functions and the associated Gauss quadrature. The main advantage of this method is its simplicity and high accuracy in the obtained energies and wave functions. With the analytical approximations for the wave functions provided by the Lagrange-mesh method allowed quadrupole transition probabilities are calculated. The CODATA 1986  fundamental constant $m_t= 5496. 921\,58\,m_e$ is used.

Because this method was applied successfully for the three body systems with Coulomb interaction H$_2^+$~\cite{OPB12}, D$_2^+$~\cite{OP13} and HD$^+$~\cite{OPB13} the reader is  referred to  these references for details. 
In Sec.~\ref{s2} only the expression for the quadruple transition is presented.  In Sec. \ref{s3}, energies are given for the lowest four vibrational levels over the full rotational bands and E2 transition probabilities are tabulated. Concluding remarks are presented in Sec.~\ref{s4}. Throughout atomic units are used. 
 
\section{Quadrupole transition probabilities}
\label{s2}

The electric quadrupole transition probability for spontaneous emission $(E_f < E_i)$ per time unit  (the atomic unit of time is $a_0/ \alpha c \approx 2.418\,8843\times 10^{-17} s$) between an initial state $i$ and a final state $f$ is given by
\begin{equation}
W_{i\rightarrow f}^{(2)}=\frac{1}{15}\alpha^{5}(E_i-E_f)^5 \frac{S_{if}^{(2)}}{2J_i+1},
\end{equation}
where $\alpha$ is the fine-structure constant, $S_{if}^{(2)}$ is the reduced matrix element and $J_i$ the total angular momentum of the initial state. The quadrupole oscillator strength is given by
\begin{equation}
f_{i\rightarrow f}^{(2)}=\frac{1}{30}\alpha^{2}(E_f-E_i)^3 \frac{S_{if}^{(2)}}{2J_i+1}.
\end{equation}
The evaluation of the reduced matrix element $S_{if}^{(2)}$ makes necessary to have the wave functions of the final and initial states. In our approximation, the wave functions are obtained using the Lagrange-mesh method.
In order to apply this method, we use the center of mass coordinates. The six-dimensional wave function is presented as a product of two three-dimensional functions, ($i$) a function carrying all the angular dependence, and ($ii$) a function  describing the form of the triangle formed by the three particles. For this internal degrees of freedom, the function is expand in a  three-dimensional Lagrange functions which are noting but a product of three one-dimension Lagrange functions. The size of each one-dimensional bases are indicated by $N_x$, $N_y$ and $N_z$. Because of the symmetry between the two centers $N_x = N_y \equiv N$. Three additional  parameters $h_x$, $h_y$ and $h_z$ are introduced in order to adjust the base to the physical problem. Due to the symmetry between the two centers $h_x=h_y\equiv h$. Any state is labeled by the total angular momentum $L$, the vibrational quantum number $v$ and the parity $\pi$: $(L^{\pi},v)$.
For a detailed discussion see~\cite{OPB12,OP13,OPB13}.

\section{T$_2^+$ bound and quasibound energies}
\label{s3}

The rotational-vibrational  spectra of the molecular ion T$_2^+$ is obtained using the Lagrange mesh method. Energies for the four lowest vibrational states ($v=0,1,2,3$) of the ground electronic state are presented in Table~\ref{tab:1}.  An important part of this work is to obtain the quadrupole transition probability per time unit, then,  it is convenient to use a single three-dimensional mesh for all states. An excellent accuracy is obtained when the size of the bases are chosen as $N= 54$ and $N_z =18$ and  the scaling parameters as $h=0.08$ and $h_z =0.6$.  In this paper the triton mass value  $m_t = 5496.921\,58\,m_e$ is used. The  dissociation energy  is then at $E_d = -0.499\,909\,056\,5$ a.u.   The first line for  each $L$-value in Table~\ref{tab:1} presents the obtained energies. Comparison with previous results are possible only for the lowest twelve rotational states with vibrational quantum number $v=0$. The energy for the ground state  $(0^+,0)$ is well known and our result is in agreement in 13 significant digits with the most accurate results presented by Frolov~\cite{Fr02} and Yan {\it et al}~\cite{YZL03}. The first rotational state $(1^-,0)$ was studied also by Yan {\it et al}~\cite{YZL03} and the agreement is in the 13 significant digit. Energies of the excited states $(L^{\pi},0)$ for $L\in(2,12)$ are reported by Yan {\it et al}~\cite{YZ04} and all of them have the same correspondence of 13 significant digits with our results. 

\begin{figure}
\begin{center}
\includegraphics[width=12cm]{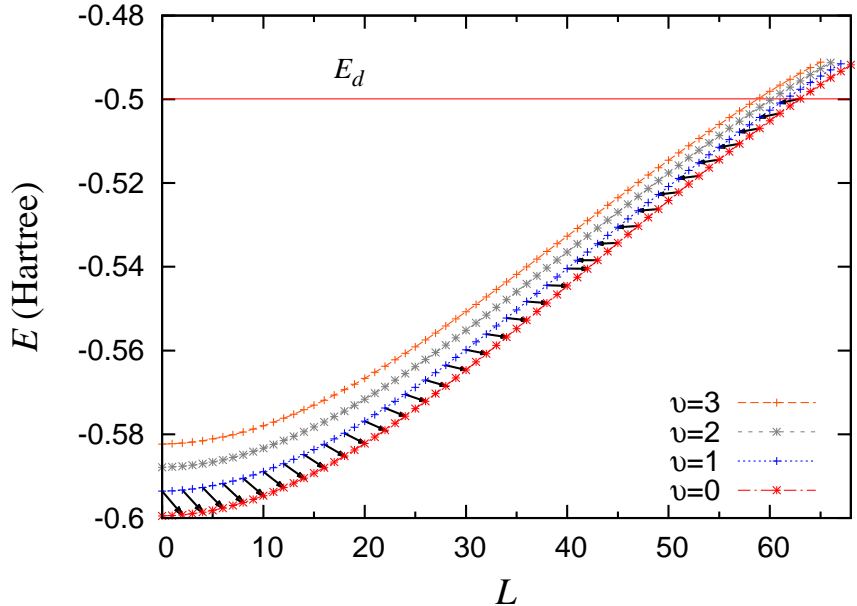}
\caption{Four lowest $\Sigma_g$ rotational bands of the T$_2^+$ molecular ion 
and dissociation energy $E_d$.  Arrows show how the direction of $L
\rightarrow L+2$  transitions between the two lowest bands changes along the band.}
\label{fig:1}
\end{center}
\end{figure}

Convergence for the energies was tested by changing the values of $N$ and $N_z$. The accuracies for the first, second and third rotational bands are $10^{-11}$, $10^{-10}$ and $10^{-9}$, respectively.
The results  demonstrate that  the vibrational band with  quantum number $v=0$ supports 62 rotational states below the dissociation energy.  The number of rotational states below the dissociation energy supported  by excited vibrational bands decrease with the vibrational quantum number: 61, 60 and 58 for  $v=1,2,3$, respectively. The obtained spectrum is depicted in Figure~\ref{fig:1}.

\begin{center}
\begin{longtable}{rllll}
\caption{Energies of the four lowest vibrational bound or quasibound states in the
$\Sigma_g$ rotational band of the T$_2^+$ molecular ion. Quasibound states are
separated from bound states by a horizontal bar.  For each $L$ value the first line
presents the Lagrange-mesh results obtained with $N_x = N_y = 54$, $N_z = 18$
and $h_x = h_y = 0.08$, $h_z = 0.6$. More
accurate energies are given for some levels (a: \cite{Fr02},
b:\cite{YZL03}, c: \cite{YZ04}). The triton mass is taken as $m_t=5496.92158\,m_e$.}
\label{tab:1}\\
\hline 
$L$& $v=0$          & $v=1$           &  $v=2$          &$v=3$\\
\hline
\endfirsthead
\multicolumn{5}{c}{{\tablename} \thetable{} -- Continuation}\\ 
\hline
$L$& $v=0$          & $v=1$           &  $v=2$          &$v=3$\\
\hline
\endhead
\hline
\multicolumn{5}{l}{{Continued on Next Page\ldots}}\\
\endfoot
\hline                                       
\endlastfoot
 0 &-0.599\,506\,910\,111\,5 &-0.593\,589\,927\,812 &-0.587\,871\,233\,66 &-0.582\,346\,606\,1\\
    &-0.599\,506\,910\,111\,54$^{a}$&&&\\
    &-0.599\,506\,910\,111\,541$^{b}$&&&\\
 1 &-0.599\,417\,152\,359\,8 &-0.593\,502\,913\,068 &-0.587\,786\,903\,36 &-0.582\,264\,906\,3\\
    &-0.599\,417\,152\,359\,852$^{b}$&&&\\
 2 &-0.599\,237\,876\,293\,2 &-0.593\,329\,117\,147 &-0.587\,618\,470\,66 &-0.582\,101\,729\,3\\
    &-0.599\,237\,876\,293\,205$^{c}$&&&\\
 3 &-0.598\,969\,558\,697\,0 &-0.593\,069\,005\,140 &-0.587\,366\,389\,36 &-0.581\,857\,517\,8\\
    &-0.598\,969\,558\,696\,973$^{c}$&&&\\
 4 &-0.598\,612\,909\,571\,6 &-0.592\,723\,269\,598 &-0.587\,031\,335\,15 &-0.581\,532\,931\,3\\
    &-0.598\,612\,909\,571\,611$^{c}$&&&\\
 5 &-0.598\,168\,866\,037\,2 &-0.592\,292\,824\,563 &-0.586\,614\,199\,81 &-0.581\,128\,840\,1\\
    &-0.598\,168\,866\,037\,205$^{c}$&&&\\
 6 &-0.597\,638\,584\,397\,7 &-0.591\,778\,797\,787 &-0.586\,116\,083\,51 &-0.580\,646\,317\,7\\
    &-0.597\,638\,584\,397\,716$^{c}$&&&\\
 7 &-0.597\,023\,430\,520\,0 &-0.591\,182\,521\,300 &-0.585\,538\,285\,67 &-0.580\,086\,632\,0\\
    &-0.597\,023\,430\,520\,051$^{c}$&&&\\
 8 &-0.596\,324\,968\,711\,3 &-0.590\,505\,520\,522 &-0.584\,882\,294\,18 &-0.579\,451\,235\,0\\
    &-0.596\,324\,968\,711\,257$^{c}$&&&\\
 9 &-0.595\,544\,949\,299\,1 &-0.589\,749\,502\,104 &-0.584\,149\,773\,62 &-0.578\,741\,750\,4\\
    &-0.595\,544\,949\,299\,086$^{c}$&&&\\
10 &-0.594\,685\,295\,136\,5 &-0.588\,916\,340\,727 &-0.583\,342\,552\,27 &-0.577\,959\,962\,1\\
     &-0.594\,685\,295\,136\,511$^{c}$&&&\\
11 &-0.593\,748\,087\,259\,4 &-0.588\,008\,065\,087 &-0.582\,462\,608\,44 &-0.577\,107\,799\,7\\
     &-0.593\,748\,087\,259\,376$^{c}$&&&\\
12 &-0.592\,735\,549\,928\,4 &-0.587\,026\,843\,285 &-0.581\,512\,056\,21 &-0.576\,187\,325\,4\\
     &-0.592\,735\,549\,928\,400$^{c}$&&&\\
13 &-0.591\,650\,035\,282\,8 &-0.585\,974\,967\,849 &-0.580\,493\,130\,81 &-0.575\,200\,719\,5\\
14 &-0.590\,494\,007\,823\,0 &-0.584\,854\,840\,613 &-0.579\,408\,173\,90 &-0.574\,150\,265\,8\\
15 &-0.589\,270\,028\,927\,2 &-0.583\,668\,957\,635 &-0.578\,259\,618\,84 &-0.573\,038\,337\,6\\
16 &-0.587\,980\,741\,586\,5 &-0.582\,419\,894\,344 &-0.577\,049\,976\,31 &-0.571\,867\,383\,5\\
17 &-0.586\,628\,855\,526\,3 &-0.581\,110\,291\,083 &-0.575\,781\,820\,19 &-0.570\,639\,913\,7\\
18 &-0.585\,217\,132\,856\,0 &-0.579\,742\,839\,179 &-0.574\,457\,774\,08 &-0.569\,358\,487\,3\\
19 &-0.583\,748\,374\,370\,8 &-0.578\,320\,267\,657 &-0.573\,080\,498\,44 &-0.568\,025\,699\,1\\
20 &-0.582\,225\,406\,601\,3 &-0.576\,845\,330\,704 &-0.571\,652\,678\,40 &-0.566\,644\,168\,6\\
21 &-0.580\,651\,069\,688\,7 &-0.575\,320\,795\,939 &-0.570\,177\,012\,45 &-0.565\,216\,528\,3\\
22 &-0.579\,028\,206\,140\,4 &-0.573\,749\,433\,558 &-0.568\,656\,201\,95 &-0.563\,745\,414\,4\\
23 &-0.577\,359\,650\,502\,1 &-0.572\,134\,006\,365 &-0.567\,092\,941\,54 &-0.562\,233\,456\,7\\
24 &-0.575\,648\,219\,963\,7 &-0.570\,477\,260\,739 &-0.565\,489\,910\,42 &-0.560\,683\,270\,9\\
25 &-0.573\,896\,705\,902\,6 &-0.568\,781\,918\,497 &-0.563\,849\,764\,56 &-0.559\,097\,450\,9\\
26 &-0.572\,107\,866\,354\,7 &-0.567\,050\,669\,681 &-0.562\,175\,129\,76 &-0.557\,478\,562\,0\\
27 &-0.570\,284\,419\,389\,4 &-0.565\,286\,166\,215 &-0.560\,468\,595\,63 &-0.555\,829\,135\,2\\
28 &-0.568\,429\,037\,360\,5 &-0.563\,491\,016\,423 &-0.558\,732\,710\,35 &-0.554\,151\,662\,7\\
29 &-0.566\,544\,341\,992\,9 &-0.561\,667\,780\,362 &-0.556\,969\,976\,24 &-0.552\,448\,592\,9\\
30 &-0.564\,632\,900\,264\,8 &-0.559\,818\,965\,926 &-0.555\,182\,846\,09 &-0.550\,722\,327\,7\\
31 &-0.562\,697\,221\,036\,4 &-0.557\,947\,025\,683 &-0.553\,373\,720\,25 &-0.548\,975\,219\,3\\
32 &-0.560\,739\,752\,377\,7 &-0.556\,054\,354\,391 &-0.551\,544\,944\,28 &-0.547\,209\,568\,5\\
33 &-0.558\,762\,879\,545\,2 &-0.554\,143\,287\,149 &-0.549\,698\,807\,35 &-0.545\,427\,622\,9\\
34 &-0.556\,768\,923\,558\,7 &-0.552\,216\,098\,133 &-0.547\,837\,541\,08 &-0.543\,631\,576\,1\\
35 &-0.554\,760\,140\,328\,2 &-0.550\,274\,999\,878 &-0.545\,963\,319\,07 &-0.541\,823\,567\,5\\
36 &-0.552\,738\,720\,285\,9 &-0.548\,322\,143\,055 &-0.544\,078\,256\,80 &-0.540\,005\,682\,4\\
37 &-0.550\,706\,788\,477\,6 &-0.546\,359\,616\,697 &-0.542\,184\,412\,02 &-0.538\,179\,952\,4\\
38 &-0.548\,666\,405\,072\,7 &-0.544\,389\,448\,859 &-0.540\,283\,785\,61 &-0.536\,348\,356\,7\\
39 &-0.546\,619\,566\,254\,9 &-0.542\,413\,607\,645 &-0.538\,378\,322\,77 &-0.534\,512\,823\,4\\
40 &-0.544\,568\,205\,458\,6 &-0.540\,434\,002\,600 &-0.536\,469\,914\,65 &-0.532\,675\,231\,4\\
41 &-0.542\,514\,194\,920\,6 &-0.538\,452\,486\,430 &-0.534\,560\,400\,24 &-0.530\,837\,412\,8\\
42 &-0.540\,459\,347\,520\,8 &-0.536\,470\,857\,022 &-0.532\,651\,568\,69 &-0.529\,001\,155\,2\\
43 &-0.538\,405\,418\,889\,7 &-0.534\,490\,859\,764 &-0.530\,745\,161\,89 &-0.527\,168\,205\,1\\
44 &-0.536\,354\,109\,765\,3 &-0.532\,514\,190\,143 &-0.528\,842\,877\,46 &-0.525\,340\,271\,0\\
45 &-0.534\,307\,068\,587\,2 &-0.530\,542\,496\,620 &-0.526\,946\,371\,95 &-0.523\,519\,027\,4\\
46 &-0.532\,265\,894\,319\,7 &-0.528\,577\,383\,791 &-0.525\,057\,264\,58 &-0.521\,706\,119\,2\\
47 &-0.530\,232\,139\,504\,5 &-0.526\,620\,415\,839 &-0.523\,177\,141\,20 &-0.519\,903\,166\,6\\
48 &-0.528\,207\,313\,545\,9 &-0.524\,673\,120\,301 &-0.521\,307\,558\,80 &-0.518\,111\,770\,4\\
49 &-0.526\,192\,886\,243\,2 &-0.522\,736\,992\,180 &-0.519\,450\,050\,49 &-0.516\,333\,518\,6\\
50 &-0.524\,190\,291\,589\,6 &-0.520\,813\,498\,456 &-0.517\,606\,131\,04 &-0.514\,569\,993\,2\\
51 &-0.522\,200\,931\,870\,4 &-0.518\,904\,083\,055 &-0.515\,777\,303\,23 &-0.512\,822\,778\,9\\
52 &-0.520\,226\,182\,104\,3 &-0.517\,010\,172\,357 &-0.513\,965\,064\,96 &-0.511\,093\,472\,8\\
53 &-0.518\,267\,394\,886\,3 &-0.515\,133\,181\,376 &-0.512\,170\,917\,53 &-0.509\,383\,695\,9\\
54 &-0.516\,325\,905\,713\,6 &-0.513\,274\,520\,748 &-0.510\,396\,375\,32 &-0.507\,695\,107\,0\\
55 &-0.514\,403\,038\,899\,0 &-0.511\,435\,604\,758 &-0.508\,642\,977\,24 &-0.506\,029\,420\,6\\
56 &-0.512\,500\,114\,211\,8 &-0.509\,617\,860\,665 &-0.506\,912\,300\,64 &-0.504\,388\,428\,3\\
57 &-0.510\,618\,454\,433\,7 &-0.507\,822\,739\,742 &-0.505\,205\,978\,29 &-0.502\,774\,027\,1\\
58 &-0.508\,759\,394\,080\,4 &-0.506\,051\,730\,550 &-0.503\,525\,719\,95 &-0.501\,188\,256\,2\\\cline{5-5}
59 &-0.506\,924\,289\,631\,3 &-0.504\,306\,375\,245 &-0.501\,873\,339\,99 &-0.499\,633\,348\,3\\
60 &-0.505\,114\,531\,741\,2 &-0.502\,588\,290\,012 &-0.500\,250\,794\,05 &-0.498\,111\,800\,5\\\cline{4-4}
61 &-0.503\,331\,560\,100\,5 &-0.500\,899\,191\,343 &-0.498\,660\,229\,06 &-0.496\,626\,48\\\cline{3-3}
62 &-0.501\,576\,881\,907\,7 &-0.499\,240\,930\,722 &-0.497\,104\,054\,16 &-0.495\,180\,78\\\cline{2-2}
63 &-0.499\,852\,095\,376\,7 &-0.497\,615\,541\,903 &-0.495\,585\,045\,64 &-0.493\,778\,9\\
64 &-0.498\,158\,920\,453\,2 &-0.496\,025\,307\,747 &-0.494\,106\,511 &-0.492\,426\\
65 &-0.496\,499\,240\,184\,4 &-0.494\,472\,859\,0     &-0.492\,672\,5&-0.491\,13\\
66 &-0.494\,875\,158\,447\,9 &-0.492\,961\,329          &-0.491\,288  &\\
67 &-0.493\,289\,084\,03        &-0.491\,494\,6              & & \\
68 &-0.491\,743\,86                 &&&\\
\end{longtable}
\end{center}       

Quadrupole transitions per time unit are easily obtained using the wave functions obtained with the Lagrange-mesh method. Table~\ref{tab:2} presents all the probabilities per second for transitions within the same rotational band, $L_f = L_i - 2$ and $v_f = v_i \le 3$. Some transition probabilities involving quasibound states are presented and these are separated by a horizontal bar. The probabilities increase slowly with $L$  reaching a maximum at $L= 55, 53, 52$ and $50$ for $v=0,1,2$ and $3$, respectively.

\begin{center}
\begin{longtable}{rllll}
\caption{Quadrupole transition probabilities per second $W$ for transitions 
between states of a same rotational band ($v_f = v_i$, $L_f = L_i - 2$). 
Results are given with five digits followed by the power of 10. } 
\label{tab:2} \\
\\[-4.9ex]
\hline
$L_i$&$v_i=0$&$v_i=1$&$v_i=2$&$v_i=3$\\
\hline
\endfirsthead
\multicolumn{4}{c}{{\tablename} \thetable{} -- Continuation}\\
\hline
$L_i$&$v_i=0$&$v_i=1$&$v_i=2$&$v_i=3$\\
\hline
\endhead
\hline
\multicolumn{4}{l}{{Continued on Next Page\ldots}}\\
\endfoot
\hline
\endlastfoot
 2& 4.077\,29-14& 4.087\,46-14& 4.069\,14-14& 4.024\,12-14\\
 3& 6.703\,22-13& 6.717\,68-13& 6.685\,44-13& 6.609\,46-13\\
 4& 3.971\,67-12& 3.978\,26-12& 3.957\,28-12& 3.910\,52-12\\
 5& 1.463\,26-11& 1.464\,72-11& 1.456\,08-11& 1.438\,00-11\\
 6& 4.086\,27-11& 4.086\,99-11& 4.059\,72-11& 4.006\,31-11\\
 7& 9.510\,01-11& 9.502\,44-11& 9.430\,29-11& 9.297\,96-11\\
 8& 1.944\,69-10& 1.940\,97-10& 1.924\,18-10& 1.895\,23-10\\
 9& 3.609\,63-10& 3.598\,19-10& 3.562\,77-10& 3.505\,11-10\\
10& 6.213\,50-10& 6.185\,18-10& 6.116\,15-10& 6.009\,44-10\\
11& 1.006\,81-09& 1.000\,69-09& 9.880\,86-10& 9.694\,84-10\\
12& 1.552\,29-09& 1.540\,34-09& 1.518\,55-09& 1.487\,70-09\\
13& 2.295\,73-09& 2.274\,07-09& 2.238\,13-09& 2.189\,10-09\\
14& 3.277\,02-09& 3.240\,08-09& 3.183\,20-09& 3.108\,09-09\\
15& 4.536\,88-09& 4.477\,00-09& 4.390\,15-09& 4.278\,75-09\\
16& 6.115\,72-09& 6.022\,70-09& 5.894\,26-09& 5.733\,72-09\\
17& 8.052\,40-09& 7.913\,12-09& 7.728\,51-09& 7.503\,01-09\\
18& 1.038\,31-08& 1.018\,11-08& 9.922\,49-09& 9.612\,99-09\\
19& 1.314\,02-08& 1.285\,55-08& 1.250\,14-08& 1.208\,55-08\\
20& 1.635\,14-08& 1.595\,99-08& 1.548\,53-08& 1.493\,69-08\\
21& 2.003\,87-08& 1.951\,24-08& 1.888\,83-08& 1.817\,78-08\\
22& 2.421\,80-08& 2.352\,47-08& 2.271\,81-08& 2.181\,23-08\\
23& 2.889\,84-08& 2.800\,16-08& 2.697\,59-08& 2.583\,81-08\\
24& 3.408\,19-08& 3.294\,11-08& 3.165\,59-08& 3.024\,62-08\\
25& 3.976\,35-08& 3.833\,42-08& 3.674\,57-08& 3.502\,11-08\\
26& 4.593\,08-08& 4.416\,49-08& 4.222\,61-08& 4.014\,11-08\\
27& 5.256\,45-08& 5.041\,08-08& 4.807\,19-08& 4.557\,85-08\\
28& 5.963\,87-08& 5.704\,28-08& 5.425\,20-08& 5.130\,06-08\\
29& 6.712\,11-08& 6.402\,66-08& 6.072\,98-08& 5.726\,95-08\\
30& 7.497\,37-08& 7.132\,21-08& 6.746\,44-08& 6.344\,33-08\\
31& 8.315\,34-08& 7.888\,51-08& 7.441\,05-08& 6.977\,67-08\\
32& 9.161\,25-08& 8.666\,71-08& 8.151\,96-08& 7.622\,13-08\\
33& 1.002\,99-07& 9.461\,67-08& 8.874\,06-08& 8.272\,68-08\\
34& 1.091\,60-07& 1.026\,80-07& 9.602\,01-08& 8.924\,12-08\\
35& 1.181\,37-07& 1.108\,00-07& 1.033\,04-07& 9.571\,18-08\\
36& 1.271\,71-07& 1.189\,20-07& 1.105\,36-07& 1.020\,85-07\\
37& 1.362\,03-07& 1.269\,83-07& 1.176\,62-07& 1.083\,10-07\\
38& 1.451\,72-07& 1.349\,31-07& 1.246\,26-07& 1.143\,32-07\\
39& 1.540\,19-07& 1.427\,07-07& 1.313\,75-07& 1.201\,02-07\\
40& 1.626\,84-07& 1.502\,54-07& 1.378\,55-07& 1.255\,71-07\\
41& 1.711\,09-07& 1.575\,18-07& 1.440\,17-07& 1.306\,91-07\\
42& 1.792\,38-07& 1.644\,46-07& 1.498\,11-07& 1.354\,19-07\\
43& 1.870\,16-07& 1.709\,89-07& 1.551\,90-07& 1.397\,11-07\\
44& 1.943\,90-07& 1.770\,97-07& 1.601\,12-07& 1.435\,28-07\\
45& 2.013\,13-07& 1.827\,27-07& 1.645\,36-07& 1.468\,33-07\\
46& 2.077\,36-07& 1.878\,36-07& 1.684\,22-07& 1.495\,92-07\\
47& 2.136\,18-07& 1.923\,84-07& 1.717\,37-07& 1.517\,76-07\\
48& 2.189\,17-07& 1.963\,36-07& 1.744\,48-07& 1.533\,55-07\\
49& 2.235\,97-07& 1.996\,58-07& 1.765\,27-07& 1.543\,04-07\\
50& 2.276\,23-07& 2.023\,22-07& 1.779\,47-07& 1.546\,02-07\\
51& 2.309\,67-07& 2.042\,99-07& 1.786\,85-07& 1.542\,28-07\\
52& 2.336\,01-07& 2.055\,68-07& 1.787\,20-07& 1.531\,66-07\\
53& 2.355\,00-07& 2.061\,05-07& 1.780\,35-07& 1.513\,99-07\\
54& 2.366\,45-07& 2.058\,95-07& 1.766\,13-07& 1.489\,14-07\\
55& 2.370\,16-07& 2.049\,19-07& 1.744\,41-07& 1.456\,98-07\\
56& 2.366\,00-07& 2.031\,65-07& 1.715\,05-07& 1.417\,41-07\\
57& 2.353\,84-07& 2.006\,20-07& 1.677\,93-07& 1.370\,29-07\\
58& 2.333\,55-07& 1.972\,73-07& 1.632\,94-07& 1.315\,50-07\\
59& 2.305\,07-07& 1.931\,13-07& 1.579\,94-07& 1.252\,87-07\\
60& 2.268\,30-07& 1.881\,30-07& 1.518\,77-07& 1.182\,18-07\\
61& 2.223\,19-07& 1.823\,10-07& 1.449\,24-07& 1.103\,12-07\\
62& 2.169\,64-07& 1.756\,37-07& 1.371\,06-07& 1.015\,22-07\\
63& 2.107\,59-07& 1.680\,90-07& 1.283\,81-07& 9.177\,09-08\\
64& 2.036\,91-07& 1.596\,37-07& 1.186\,85-07& 8.093\,00-08\\
65& 1.957\,44-07& 1.502\,31-07& 1.079\,16-07& 6.874\,67-08\\
66& 1.868\,92-07& 1.397\,98-07& 9.589\,20-08& \\
67& 1.770\,96-07& 1.282\,13-07& &\\
68& 1.662\,90-07& &&\\
\end{longtable}
\end{center}

Quadrupole transitions per time unit for other transitions are presented in Table~\ref{tab:3}. The columns correspond to transitions between different vibrational states. For each $L_i$ value, the successive lines correspond to increasing values of $L_f$, i.e. to $L_f =L_i - 2$ for $L_i >1$, $L_f =L_i$ for $L_i > 0$, and $L_f =L_i+2$, respectively. The strongest transition from each state occurs in general towards the nearest vibrational state $(v_f = v_i -1)$ for $L_f = L_i - 2$. For $v_f = v_i -1$, in the vicinity of $L_i = 41$, and beyond, the $(L_i, v_i )\rightarrow (L_i + 2, vi-1)$ transitions are replaced by $(L_i + 2, vi -1) \rightarrow (L_i, v_i )$ transitions because the sign of the energy difference changes (see the arrows in Figure~\ref{fig:1} for the $1 0 $ transitions). These numbers are indicated in italics in Table~\ref{tab:3}. For example, the first number in the last line for $L_i = 41$ corresponds to the $(43, 0) \rightarrow (41, 1)$ transition. Hence, the transition probabilities strongly vary in this region.

\begin{center}
\begin{longtable}{lllllll}
\caption{Quadrupole transition probabilities per second $W$ for transitions 
between different vibrational quantum numbers $(v_i \ne v_f)$. 
The three successive lines correspond to increasing $L_f$ values, i.e.\ 
$L_f = L_i - 2$, $L_f = L_i$ and $L_f = L_i + 2$, respectively, for $L_i \ge 2$. 
Italicized numbers for $(1 \rightarrow 0)$, $(2 \rightarrow 1)$ and $(3 \rightarrow 2)$ 
mean that the initial and final states are exchanged 
(the first one is preceded in each case by a horizontal bar).} 
\label{tab:3}\\
\\[-4.9ex]
\hline
$L_i$&$(1\rightarrow\,0)$&$(2\rightarrow\,0)$&$(2\rightarrow\,1)$
     &$(3\rightarrow\,0)$&$(3\rightarrow\,1)$&$(3\rightarrow\,2)$\\
\hline
\endfirsthead
\multicolumn{7}{c}{{\tablename} \thetable{} -- Continuation}\\
\hline
$L_i$&$(1\rightarrow\,0)$&$(2\rightarrow\,0)$&$(2\rightarrow\,1)$
     &$(3\rightarrow\,0)$&$(3\rightarrow\,1)$&$(3\rightarrow\,2)$\\
\hline
\endhead
\hline
\multicolumn{7}{l}{{Continued on Next Page\ldots}}\\
\endfoot
\hline
\endlastfoot
 0& 2.352\,92$-$08& 1.966\,46$-$09& 4.209\,56$-$08& 1.218\,32$-$10& 5.449\,55$-$09& 5.631\,51$-$08\\
1 & 1.076\,05$-$08& 1.025\,38$-$09& 1.920\,69$-$08& 7.676\,43$-$11& 2.817\,57$-$09& 2.563\,57$-$08\\
  & 1.278\,01$-$08& 9.691\,38$-$10& 2.289\,40$-$08& 5.114\,59$-$11& 2.702\,85$-$09& 3.066\,63$-$08\\
 2& 6.053\,60$-$09& 6.477\,38$-$10& 1.077\,63$-$08& 5.616\,33$-$11& 1.766\,56$-$09& 1.434\,44$-$08\\
  & 7.668\,00$-$09& 7.318\,48$-$10& 1.368\,58$-$08& 5.487\,71$-$11& 2.010\,80$-$09& 1.826\,48$-$08\\
  & 9.839\,31$-$09& 6.702\,09$-$10& 1.764\,53$-$08& 2.889\,74$-$11& 1.882\,55$-$09& 2.366\,10$-$08\\
 3& 8.334\,47$-$09& 9.577\,09$-$10& 1.480\,60$-$08& 9.016\,18$-$11& 2.599\,81$-$09& 1.966\,74$-$08\\
  & 7.131\,55$-$09& 6.822\,58$-$10& 1.272\,66$-$08& 5.128\,11$-$11& 1.874\,29$-$09& 1.698\,23$-$08\\
  & 8.118\,99$-$09& 4.906\,61$-$10& 1.457\,34$-$08& 1.621\,87$-$11& 1.389\,52$-$09& 1.955\,89$-$08\\
 4& 9.839\,96$-$09& 1.209\,31$-$09& 1.744\,09$-$08& 1.224\,32$-$10& 3.268\,27$-$09& 2.311\,43$-$08\\
  & 6.913\,64$-$09& 6.634\,88$-$10& 1.233\,56$-$08& 5.002\,82$-$11& 1.822\,39$-$09& 1.645\,73$-$08\\
  & 6.852\,13$-$09& 3.618\,13$-$10& 1.230\,83$-$08& 8.249\,22$-$12& 1.034\,47$-$09& 1.653\,02$-$08\\
 5& 1.100\,38$-$08& 1.441\,38$-$09& 1.945\,56$-$08& 1.557\,27$-$10& 3.878\,88$-$09& 2.571\,96$-$08\\
  & 6.784\,85$-$09& 6.536\,72$-$10& 1.210\,31$-$08& 4.948\,10$-$11& 1.795\,02$-$09& 1.614\,33$-$08\\
  & 5.825\,36$-$09& 2.636\,09$-$10& 1.046\,95$-$08& 3.389\,76$-$12& 7.623\,79$-$10& 1.406\,77$-$08\\
 6& 1.195\,55$-$08& 1.664\,33$-$09& 2.108\,16$-$08& 1.906\,95$-$10& 4.460\,38$-$09& 2.779\,29$-$08\\
  & 6.687\,94$-$09& 6.473\,36$-$10& 1.192\,71$-$08& 4.922\,88$-$11& 1.777\,14$-$09& 1.590\,39$-$08\\
  & 4.956\,19$-$09& 1.874\,69$-$10& 8.910\,54$-$09& 8.223\,10$-$13& 5.499\,08$-$10& 1.197\,65$-$08\\
 7& 1.274\,29$-$08& 1.880\,78$-$09& 2.240\,48$-$08& 2.273\,62$-$10& 5.020\,13$-$09& 2.944\,95$-$08\\
  & 6.602\,78$-$09& 6.425\,42$-$10& 1.177\,15$-$08& 4.912\,48$-$11& 1.763\,42$-$09& 1.569\,13$-$08\\
  & 4.205\,83$-$09& 1.286\,50$-$10& 7.562\,74$-$09& 3.163$-$15      & 3.842\,83$-$10& 1.016\,60$-$08\\
 8& 1.338\,44$-$08& 2.090\,35$-$09& 2.345\,90$-$08& 2.654\,77$-$10& 5.557\,41$-$09& 3.073\,59$-$08\\
  & 6.520\,84$-$09& 6.384\,48$-$10& 1.162\,13$-$08& 4.910\,45$-$11& 1.751\,56$-$09& 1.548\,51$-$08\\
  & 3.553\,06$-$09& 8.398\,43$-$11& 6.388\,81$-$09& 5.151\,4 $-$13& 2.570\,35$-$10& 8.587\,19$-$09\\
 9& 1.388\,65$-$08& 2.291\,35$-$09& 2.425\,70$-$08& 3.046\,33$-$10& 6.067\,94$-$09& 3.167\,10$-$08\\
  & 6.438\,02$-$09& 6.346\,37$-$10& 1.146\,91$-$08& 4.913\,45$-$11& 1.740\,42$-$09& 1.527\,57$-$08\\
  & 2.984\,49$-$09& 5.105\,00$-$11& 5.365\,33$-$09& 2.015\,57$-$12& 1.617\,32$-$10& 7.209\,46$-$09\\
10& 1.425\,08$-$08& 2.481\,48$-$09& 2.480\,33$-$08& 3.443\,26$-$10& 6.545\,88$-$09& 3.226\,30$-$08\\
  & 6.352\,26$-$09& 6.308\,78$-$10& 1.131\,13$-$08& 4.919\,57$-$11& 1.729\,36$-$09& 1.505\,82$-$08\\
  & 2.490\,52$-$09& 2.784\,36$-$11& 4.475\,49$-$09& 4.215\,75$-$12& 9.305\,25$-$11& 6.010\,82$-$09\\
11& 1.447\,76$-$08& 2.658\,16$-$09& 2.510\,01$-$08& 3.839\,88$-$10& 6.984\,74$-$09& 3.251\,72$-$08\\
  & 6.262\,43$-$09& 6.270\,34$-$10& 1.114\,59$-$08& 4.927\,55$-$11& 1.718\,00$-$09& 1.483\,01$-$08\\
  & 2.063\,39$-$09& 1.263\,30$-$11& 3.705\,70$-$09& 6.871\,68$-$12& 4.639\,12$-$11& 4.973\,47$-$09\\
12& 1.456\,72$-$08& 2.818\,84$-$09& 2.515\,01$-$08& 4.230\,17$-$10& 7.378\,14$-$09& 3.244\,01$-$08\\
  & 6.167\,98$-$09& 6.230\,15$-$10& 1.097\,19$-$08& 4.936\,50$-$11& 1.706\,11$-$09& 1.459\,01$-$08\\
  & 1.696\,33$-$09& 3.893\,88$-$12& 3.044\,04$-$09& 9.779\,27$-$12& 1.768\,15$-$11& 4.081\,72$-$09\\
13& 1.452\,14$-$08& 2.961\,08$-$09& 2.495\,86$-$08& 4.608\,01$-$10& 7.720\,13$-$09& 3.204\,18$-$08\\
  & 6.068\,67$-$09& 6.187\,63$-$10& 1.078\,89$-$08& 4.945\,75$-$11& 1.693\,54$-$09& 1.433\,77$-$08\\
  & 1.383\,15$-$09& 2.803\,80$-$13& 2.479\,52$-$09& 1.277\,08$-$11& 3.317\,84$-$12& 3.320\,98$-$09\\
14& 1.434\,34$-$08& 3.082\,73$-$09& 2.453\,38$-$08& 4.967\,37$-$10& 8.005\,53$-$09& 3.133\,67$-$08\\
  & 5.964\,42$-$09& 6.142\,34$-$10& 1.059\,69$-$08& 4.954\,74$-$11& 1.680\,16$-$09& 1.407\,28$-$08\\
  & 1.118\,03$-$09& 6.106\,10$-$13& 2.001\,77$-$09& 1.571\,20$-$11& 1.196\,23$-$13& 2.677\,42$-$09\\
15& 1.403\,87$-$08& 3.181\,97$-$09& 2.388\,79$-$08& 5.302\,5$-$10& 8.230\,1$-$09& 3.034\,38$-$08\\
  & 5.855\,30$-$09& 6.093\,99$-$10& 1.039\,60$-$08& 4.963\,01$-$11& 1.665\,90$-$09& 1.379\,58$-$08\\
  & 8.954\,75$-$10& 3.857\,94$-$12& 1.600\,94$-$09& 1.849\,87$-$11& 5.310\,1$-$12& 2.137\,81$-$09\\
16& 1.361\,45$-$08& 3.257\,39$-$09& 2.303\,62$-$08& 5.608\,17$-$10& 8.390\,8$-$09& 2.908\,72$-$08\\
  & 5.741\,47$-$09& 6.042\,34$-$10& 1.018\,65$-$08& 4.970\,17$-$11& 1.650\,69$-$09& 1.350\,70$-$08\\
  & 7.103\,10$-$10& 9.143\,31$-$12& 1.267\,71$-$09& 2.105\,36$-$11& 1.649\,90$-$11& 1.689\,62$-$09\\
17& 1.308\,04$-$08& 3.308\,04$-$09& 2.199\,77$-$08& 5.879\,7$-$10& 8.485\,83$-$09& 2.759\,50$-$08\\
  & 5.623\,14$-$09& 5.987\,25$-$10& 9.968\,78$-$09& 4.975\,87$-$11& 1.634\,51$-$09& 1.320\,71$-$08\\
  & 5.576\,85$-$10& 1.572\,69$-$11& 9.933\,25$-$10& 2.332\,30$-$11& 3.166\,31$-$11& 1.321\,02$-$09\\
18& 1.244\,73$-$08& 3.333\,43$-$09& 2.079\,44$-$08& 6.113\,1$-$10& 8.514\,4$-$09& 2.589\,92$-$08\\
  & 5.500\,56$-$09& 5.928\,59$-$10& 9.743\,38$-$09& 4.979\,82$-$11& 1.617\,32$-$09& 1.289\,68$-$08\\
  & 4.331\,11$-$10& 2.299\,86$-$11& 7.696\,66$-$10& 2.527\,31$-$11& 4.912\,29$-$11& 1.021\,00$-$09\\
19& 1.172\,80$-$08& 3.333\,5$-$09& 1.945\,07$-$08& 6.305\,3$-$10& 8.477\,1$-$09& 2.403\,50$-$08\\
  & 5.374\,02$-$09& 5.866\,29$-$10& 9.510\,84$-$09& 4.981\,75$-$11& 1.599\,11$-$09& 1.257\,68$-$08\\
  & 3.324\,75$-$10& 3.046\,69$-$11& 5.892\,76$-$10& 2.688\,68$-$11& 6.751\,52$-$11& 7.794\,50$-$10\\
20& 1.093\,63$-$08& 3.308\,75$-$09& 1.799\,29$-$08& 6.453\,9$-$10& 8.375\,8$-$09& 2.203\,97$-$08\\
  & 5.243\,83$-$09& 5.800\,32$-$10& 9.271\,75$-$09& 4.981\,44$-$11& 1.579\,86$-$09& 1.224\,81$-$08\\
  & 2.520\,58$-$10& 3.774\,70$-$11& 4.454\,05$-$10& 2.816\,09$-$11& 8.576\,37$-$11& 5.872\,12$-$10\\
21& 1.008\,72$-$08& 3.259\,9$-$09& 1.644\,9$-$08& 6.557\,60$-$10& 8.213\,3$-$09& 1.995\,20$-$08\\
  & 5.110\,33$-$09& 5.730\,67$-$10& 9.026\,73$-$09& 4.978\,72$-$11& 1.559\,59$-$09& 1.191\,14$-$08\\
  & 1.885\,36$-$10& 4.454\,87$-$11& 3.320\,16$-$10& 2.910\,27$-$11& 1.030\,46$-$10& 4.360\,81$-$10\\
22& 9.196\,0$-$09& 3.188\,4$-$09& 1.484\,74$-$08& 6.615\,5$-$10& 7.993\,4$-$09& 1.781\,11$-$08\\
  & 4.973\,86$-$09& 5.657\,36$-$10& 8.776\,43$-$09& 4.973\,42$-$11& 1.538\,29$-$09& 1.156\,77$-$08\\
  & 1.389\,74$-$10& 5.066\,35$-$11& 2.437\,80$-$10& 2.972\,75$-$11& 1.187\,62$-$10& 3.188\,17$-$10\\
23& 8.278\,55$-$09& 3.095\,7$-$09& 1.321\,70$-$08& 6.628\,1$-$10& 7.720\,7$-$09& 1.565\,62$-$08\\
  & 4.834\,76$-$09& 5.580\,44$-$10& 8.521\,50$-$09& 4.965\,44$-$11& 1.515\,99$-$09& 1.121\,79$-$08\\
  & 1.008\,13$-$10& 5.595\,26$-$11& 1.760\,51$-$10& 3.005\,69$-$11& 1.325\,01$-$10& 2.291\,12$-$10\\
24& 7.350\,4$-$09& 2.983\,6$-$09& 1.158\,6$-$08& 6.596\,1$-$10& 7.400\,6$-$09& 1.352\,5$-$08\\
  & 4.693\,41$-$09& 5.499\,98$-$10& 8.262\,61$-$09& 4.954\,69$-$11& 1.492\,71$-$09& 1.086\,30$-$08\\
  & 7.185\,10$-$11& 6.033\,52$-$11& 1.248\,31$-$10& 3.011\,60$-$11& 1.440\,11$-$10& 1.615\,38$-$10\\
25& 6.427\,0$-$09& 2.854\,39$-$09& 9.981\,4$-$09& 6.521\,4$-$10& 7.038\,9$-$09& 1.145\,4$-$08\\
  & 4.550\,15$-$09& 5.416\,06$-$10& 8.000\,44$-$09& 4.941\,11$-$11& 1.468\,46$-$09& 1.050\,38$-$08\\
  & 5.021\,28$-$11& 6.377\,79$-$11& 8.672\,25$-$11& 2.993\,26$-$11& 1.531\,71$-$10& 1.114\,92$-$10\\
26& 5.523\,16$-$09& 2.710\,2$-$09& 8.429\,80$-$09& 6.406\,0$-$10& 6.641\,95$-$09& 9.477\,97$-$09\\
  & 4.405\,34$-$09& 5.328\,79$-$10& 7.735\,64$-$09& 4.924\,68$-$11& 1.443\,29$-$09& 1.014\,14$-$08\\
  & 3.432\,55$-$11& 6.628\,48$-$11& 5.887\,73$-$11& 2.953\,53$-$11& 1.599\,65$-$10& 7.512\,05$-$11\\
27& 4.652\,76$-$09& 2.553\,44$-$09& 6.955\,2$-$09& 6.252\,7$-$10& 6.216\,2$-$09& 7.627\,4$-$09\\
  & 4.259\,35$-$09& 5.238\,29$-$10& 7.468\,89$-$09& 4.905\,39$-$11& 1.417\,23$-$09& 9.776\,61$-$09\\
  & 2.288\,51$-$11& 6.788\,90$-$11& 3.893\,97$-$11& 2.895\,29$-$11& 1.644\,55$-$10& 4.924\,11$-$11\\
28& 3.828\,79$-$09& 2.386\,6$-$09& 5.579\,9$-$09& 6.064\,9$-$10& 5.768\,3$-$09& 5.930\,9$-$09\\
  & 4.112\,52$-$09& 5.144\,70$-$10& 7.200\,83$-$09& 4.883\,29$-$11& 1.390\,30$-$09& 9.410\,36$-$09\\
  & 1.482\,59$-$11& 6.864\,49$-$11& 2.498\,85$-$11& 2.821\,34$-$11& 1.667\,68$-$10& 3.126\,69$-$11\\
29& 3.063\,0$-$09& 2.212\,26$-$09& 4.323\,80$-$09& 5.845\,8$-$10& 5.304\,95$-$09& 4.413\,36$-$09\\
  & 3.965\,19$-$09& 5.048\,16$-$10& 6.932\,10$-$09& 4.858\,40$-$11& 1.362\,57$-$09& 9.043\,54$-$09\\
  & 9.289\,68$-$12& 6.862\,23$-$11& 1.548\,11$-$11& 2.734\,34$-$11& 1.670\,72$-$10& 1.912\,68$-$11\\
30& 2.365\,97$-$09& 2.032\,79$-$09& 3.204\,24$-$09& 5.599\,29$-$10& 4.832\,60$-$09& 3.096\,05$-$09\\
  & 3.817\,71$-$09& 4.948\,82$-$10& 6.663\,32$-$09& 4.830\,79$-$11& 1.334\,06$-$09& 8.677\,02$-$09\\
  & 5.596\,29$-$12& 6.790\,03$-$11& 9.199\,19$-$12& 2.636\,76$-$11& 1.655\,62$-$10& 1.119\,16$-$11\\
31& 1.746\,77$-$09& 1.850\,63$-$09& 2.236\,07$-$09& 5.329\,18$-$10& 4.357\,58$-$09& 1.996\,57$-$09\\
  & 3.670\,39$-$09& 4.846\,86$-$10& 6.395\,09$-$09& 4.800\,55$-$11& 1.304\,83$-$09& 8.311\,61$-$09\\
  & 3.216\,00$-$12& 6.656\,34$-$11& 5.198\,05$-$12& 2.530\,91$-$11& 1.624\,53$-$10& 6.204\,20$-$12\\
32& 1.213\,15$-$09& 1.668\,10$-$09& 1.431\,46$-$09& 5.039\,43$-$10& 3.885\,89$-$09& 1.128\,80$-$09\\
  & 3.523\,55$-$09& 4.742\,43$-$10& 6.127\,98$-$09& 4.767\,77$-$11& 1.274\,92$-$09& 7.948\,14$-$09\\
  & 1.744\,50$-$12& 6.469\,81$-$11& 2.760\,53$-$12& 2.418\,85$-$11& 1.579\,65$-$10& 3.216\,05$-$12\\
33& 7.714\,1$-$10& 1.487\,39$-$09& 7.999\,1$-$10& 4.733\,98$-$10& 3.423\,18$-$09& 5.029\,3$-$10\\
  & 3.377\,49$-$09& 4.635\,71$-$10& 5.862\,57$-$09& 4.732\,57$-$11& 1.244\,37$-$09& 7.587\,36$-$09\\
  & 8.802\,65$-$13& 6.238\,95$-$11& 1.355\,38$-$12& 2.302\,43$-$11& 1.523\,18$-$10& 1.529\,98$-$12\\
34& 4.264\,6$-$10& 1.310\,55$-$09& 3.483\,6$-$10& 4.416\,7$-$10& 2.974\,70$-$09& 1.255\,9$-$10\\
  & 3.232\,50$-$09& 4.526\,88$-$10& 5.599\,37$-$09& 4.695\,05$-$11& 1.213\,25$-$09& 7.230\,01$-$09\\
  & 4.045\,46$-$13& 5.971\,99$-$11& 6.005\,87$-$13& 2.183\,29$-$11& 1.457\,23$-$10& 6.496\,11$-$13\\
35& 1.817\,9$-$10& 1.139\,50$-$09& 8.119$-$11& 4.091\,4$-$10& 2.545\,26$-$09& 5.494$-$16\\
  & 3.088\,87$-$09& 4.416\,10$-$10& 5.338\,90$-$09& 4.655\,34$-$11& 1.181\,59$-$09& 6.876\,79$-$09\\
  & 1.639\,80$-$13& 5.676\,68$-$11& 2.313\,56$-$13& 2.062\,86$-$11& 1.383\,84$-$10& 2.354\,82$-$13\\
36& 3.960\,1$-$11& 9.759\,5$-$10& 4.209$-$13& 3.761\,56$-$10& 2.139\,20$-$09& 1.261\,19$-$10\\
  & 2.946\,83$-$09& 4.303\,56$-$10& 5.081\,64$-$09& 4.613\,58$-$11& 1.149\,44$-$09& 6.528\,36$-$09\\
  & 5.563\,95$-$14& 5.360\,21$-$11& 7.277\,03$-$14& 1.942\,37$-$11& 1.304\,86$-$10& 6.748\,60$-$14\\
37& 8.078$-$13& 8.214\,7$-$10& 1.058\,0$-$10& 3.430\,7$-$10& 1.760\,4$-$09& 5.009\,0$-$10\\
  & 2.806\,65$-$09& 4.189\,43$-$10& 4.828\,03$-$09& 4.569\,88$-$11& 1.116\,85$-$09& 6.185\,35$-$09\\
  & 1.438\,79$-$14& 5.029\,16$-$11& 1.661\,93$-$14& 1.822\,85$-$11& 1.222\,03$-$10& 1.313\,86$-$14\\
38& 6.516$-$11& 6.774\,36$-$10& 3.950\,1$-$10& 3.102\,08$-$10& 1.412\,30$-$09& 1.118\,53$-$09\\
  & 2.668\,56$-$09& 4.073\,87$-$10& 4.578\,50$-$09& 4.524\,39$-$11& 1.083\,87$-$09& 5.848\,33$-$09\\
  & 2.331\,32$-$15& 4.689\,41$-$11& 2.118\,82$-$15& 1.705\,19$-$11& 1.136\,88$-$10& 1.197\,40$-$15\\
39& 2.313\,39$-$10& 5.450\,6$-$10& 8.637\,9$-$10& 2.778\,6$-$10& 1.097\,81$-$09& 1.970\,66$-$09\\
  & 2.532\,76$-$09& 3.957\,06$-$10& 4.333\,45$-$09& 4.477\,23$-$11& 1.050\,54$-$09& 5.517\,86$-$09\\
  & 1.386\,40$-$16& 4.346\,17$-$11& 6.554\,54$-$17& 1.590\,12$-$11& 1.050\,80$-$10& 1.160\,54$-$17\\
40& 4.970\,1$-$10& 4.253\,9$-$10& 1.506\,18$-$09& 2.463\,07$-$10& 8.194\,6$-$10& 3.046\,6$-$09\\
  & 2.399\,47$-$09& 3.839\,15$-$10& 4.093\,24$-$09& 4.428\,52$-$11& 1.016\,91$-$09& 5.194\,44$-$09\\\cline{7-7}
  & 1.484\,82$-$19& 4.004\,02$-$11& 2.290\,73$-$26& 1.478\,21$-$11& 9.649\,65$-$11& {\it 3.553\,47$-$19}\\
41& 8.589\,7$-$10& 3.193\,17$-$10& 2.314\,67$-$09& 2.158\,00$-$10& 5.793\,48$-$10& 4.333\,86$-$09\\
  & 2.268\,86$-$09& 3.720\,31$-$10& 3.858\,22$-$09& 4.378\,38$-$11& 9.830\,22$-$10& 4.878\,53$-$09\\\cline{2-2}\cline{4-4}
  & {\it 3.300\,74$-$18}& 3.666\,84$-$11& {\it 5.047\,82$-$17}& 1.369\,94$-$11& 8.804\,22$-$11& {\it 3.384\,62$-$16}\\
42& 1.313\,23$-$09& 2.275\,68$-$10& 3.280\,37$-$09& 1.865\,7$-$10& 3.791\,8$-$10& 5.817\,89$-$09\\
  & 2.141\,12$-$09& 3.600\,68$-$10& 3.628\,71$-$09& 4.326\,93$-$11& 9.489\,15$-$10& 4.570\,56$-$09\\
  & {\it 3.274\,43$-$16}& 3.337\,97$-$11& {\it 1.605\,87$-$15}& 1.265\,68$-$11& 7.980\,38$-$11& {\it 5.325\,31$-$15}\\
43& 1.855\,11$-$09& 1.507\,49$-$10& 4.393\,24$-$09& 1.588\,36$-$10& 2.203\,26$-$10& 7.482\,77$-$09\\
  & 2.016\,39$-$09& 3.480\,41$-$10& 3.404\,97$-$09& 4.274\,27$-$11& 9.146\,32$-$10& 4.270\,93$-$09\\
  & {\it 3.346\,80$-$15}& 3.020\,12$-$11& {\it 1.187\,03$-$14}& 1.165\,67$-$11& 7.185\,34$-$11& {\it 3.049\,73$-$14}\\
44& 2.479\,36$-$09& 8.933\,87$-$11& 5.642\,2$-$09& 1.327\,8$-$10& 1.038\,2$-$10& 9.311\,25$-$09\\
  & 1.894\,81$-$09& 3.359\,63$-$10& 3.187\,29$-$09& 4.220\,48$-$11& 8.802\,10$-$10& 3.980\,00$-$09\\
  & {\it 1.592\,91$-$14}& 2.715\,53$-$11& {\it 4.847\,99$-$14}& 1.070\,13$-$11& 6.424\,96$-$11& {\it 1.092\,26$-$13}\\
45& 3.180\,20$-$09& 4.370\,6$-$11& 7.015\,36$-$09& 1.085\,86$-$10& 3.043$-$11& 1.128\,50$-$08\\
  & 1.776\,53$-$09& 3.238\,48$-$10& 2.975\,88$-$09& 4.165\,65$-$11& 8.456\,87$-$10& 3.698\,07$-$09\\
  & {\it 5.143\,45$-$14}& 2.425\,93$-$11& {\it 1.432\,61$-$13}& 9.791\,66$-$12& 5.703\,82$-$11& {\it 2.980\,88$-$13}\\
46& 3.951\,50$-$09& 1.412\,8$-$11& 8.500\,1$-$09& 8.641\,2$-$11& 6.898$-$13& 1.338\,4$-$08\\
  & 1.661\,64$-$09& 3.117\,09$-$10& 2.770\,96$-$09& 4.109\,84$-$11& 8.110\,97$-$10& 3.425\,44$-$09\\
  & {\it 1.313\,00$-$13}& 2.152\,65$-$11& {\it 3.453\,00$-$13}& 8.928\,43$-$12& 5.025\,38$-$11& {\it 6.814\,32$-$13}\\
47& 4.786\,76$-$09& 8.061$-$13& 1.008\,33$-$08& 6.641\,4$-$11& 1.493\,2$-$11& 1.559\,01$-$08\\
  & 1.550\,27$-$09& 2.995\,57$-$10& 2.572\,72$-$09& 4.053\,08$-$11& 7.764\,74$-$10& 3.162\,36$-$09\\
  & {\it 2.863\,53$-$13}& 1.896\,63$-$11& {\it 7.236\,25$-$13}& 8.111\,81$-$12& 4.392\,10$-$11& {\it 1.375\,59$-$12}\\
48& 5.679\,28$-$09& 3.879\,6$-$12& 1.175\,13$-$08& 4.873\,7$-$11& 7.334\,1$-$11& 1.788\,08$-$08\\
  & 1.442\,48$-$09& 2.874\,03$-$10& 2.381\,32$-$09& 3.995\,41$-$11& 7.418\,49$-$10& 2.909\,04$-$09\\
  & {\it 5.580\,06$-$13}& 1.658\,46$-$11& {\it 1.369\,78$-$12}& 7.341\,58$-$12& 3.805\,53$-$11& {\it 2.533\,21$-$12}\\
49& 6.622\,16$-$09& 2.344\,7$-$11& 1.349\,02$-$08& 3.352\,95$-$11& 1.759\,97$-$10& 2.023\,55$-$08\\
  & 1.338\,38$-$09& 2.752\,57$-$10& 2.196\,90$-$09& 3.936\,83$-$11& 7.072\,52$-$10& 2.665\,69$-$09\\
  & {\it 9.994\,50$-$13}& 1.438\,48$-$11& {\it 2.400\,60$-$12}& 6.617\,19$-$12& 3.266\,46$-$11& {\it 4.348\,19$-$12}\\
50& 7.608\,4$-$09& 5.958$-$11& 1.528\,6$-$08& 2.094$-$11& 3.229\,1$-$10& 2.263\,2$-$08\\
  & 1.238\,01$-$09& 2.631\,31$-$10& 2.019\,58$-$09& 3.877\,32$-$11& 6.727\,13$-$10& 2.432\,45$-$09\\
  & {\it 1.677\,03$-$12}& 1.236\,73$-$11& {\it 3.961\,54$-$12}& 5.937\,86$-$12& 2.775\,01$-$11& {\it 7.061\,89$-$12}\\
51& 8.630\,9$-$09& 1.123\,5$-$10& 1.712\,3$-$08& 1.113$-$11& 5.140\,9$-$10& 2.504\,97$-$08\\
  & 1.141\,43$-$09& 2.510\,32$-$10& 1.849\,46$-$09& 3.816\,82$-$11& 6.382\,56$-$10& 2.209\,47$-$09\\
  & {\it 2.671\,95$-$12}& 1.053\,07$-$11& {\it 6.230\,94$-$12}& 5.302\,60$-$12& 2.330\,70$-$11& {\it 1.097\,12$-$11}\\
52& 9.682\,7$-$09& 1.818\,6$-$10& 1.898\,9$-$08& 4.278$-$12& 7.495\,6$-$10& 2.746\,45$-$08\\
  & 1.048\,70$-$09& 2.389\,69$-$10& 1.686\,62$-$09& 3.755\,27$-$11& 6.039\,07$-$10& 1.996\,84$-$09\\
  & {\it 4.082\,49$-$12}& 8.871\,63$-$12& {\it 9.425\,75$-$12}& 4.710\,29$-$12& 1.932\,59$-$11& {\it 1.643\,99$-$11}\\
53& 1.075\,67$-$08& 2.682\,1$-$10& 2.086\,92$-$08& 5.923$-$13& 1.029\,43$-$09& 2.985\,41$-$08\\
  & 9.598\,52$-$10& 2.269\,50$-$10& 1.531\,12$-$09& 3.692\,52$-$11& 5.696\,88$-$10& 1.794\,65$-$09\\
  & {\it 6.027\,05$-$12}& 7.385\,21$-$12& {\it 1.380\,93$-$11}& 4.159\,68$-$12& 1.579\,30$-$11& {\it 2.391\,37$-$11}\\
54& 1.184\,60$-$08& 3.716\,0$-$10& 2.274\,72$-$08& 3.184$-$13& 1.353\,9$-$09& 3.219\,47$-$08\\
  & 8.749\,05$-$10& 2.149\,82$-$10& 1.383\,02$-$09& 3.628\,41$-$11& 5.356\,22$-$10& 1.602\,97$-$09\\
  & {\it 8.648\,33$-$12}& 6.065\,34$-$12& {\it 1.970\,25$-$11}& 3.649\,46$-$12& 1.269\,14$-$11& {\it 3.394\,27$-$11}\\
55& 1.294\,36$-$08& 4.923\,0$-$10& 2.460\,91$-$08& 3.757\,7$-$12& 1.723\,58$-$09& 3.446\,19$-$08\\
  & 7.938\,82$-$10& 2.030\,72$-$10& 1.242\,34$-$09& 3.562\,72$-$11& 5.017\,25$-$10& 1.421\,82$-$09\\
  & {\it 1.211\,91$-$11}& 4.904\,85$-$12& {\it 2.749\,85$-$11}& 3.178\,28$-$12& 1.000\,12$-$11& {\it 4.721\,21$-$11}\\
56& 1.404\,29$-$08& 6.307\,32$-$10& 2.643\,99$-$08& 1.128\,6$-$11& 2.139\,01$-$09& 3.663\,00$-$08\\
  & 7.167\,94$-$10& 1.912\,25$-$10& 1.109\,10$-$09& 3.495\,14$-$11& 4.680\,16$-$10& 1.251\,22$-$09\\
  & {\it 1.665\,01$-$1}1& 3.895\,79$-$12& {\it 3.768\,54$-$11}& 2.744\,75$-$12& 7.700\,17$-$12& {\it 6.458\,86$-$11}\\
57& 1.513\,72$-$08& 7.874\,6$-$10& 2.822\,44$-$08& 2.337\,9$-$11& 2.601\,26$-$09& 3.867\,15$-$08\\
  & 6.436\,44$-$10& 1.794\,46$-$10& 9.832\,96$-$10& 3.425\,27$-$11& 4.345\,05$-$10& 1.091\,18$-$09\\
  & {\it 2.250\,18$-$11}& 3.029\,52$-$12& {\it 5.087\,71$-$11}& 2.347\,51$-$12& 5.764\,32$-$12& {\it 8.718\,78$-$11}\\
58& 1.622\,00$-$08& 9.632\,6$-$10& 2.994\,65$-$08& 4.064\,8$-$11& 3.111\,74$-$09& 4.055\,65$-$08\\
  & 5.744\,28$-$10& 1.677\,39$-$10& 8.649\,25$-$10& 3.352\,61$-$11& 4.012\,03$-$10& 9.416\,67$-$10\\
  & {\it 3.000\,06$-$11}& 2.296\,89$-$12& {\it 6.786\,01$-$11}& 1.985\,18$-$12& 4.167\,99$-$12& {\it 1.164\,78$-$10}\\
59& 1.728\,49$-$08& 1.159\,21$-$09& 3.158\,95$-$08& 6.389\,1$-$11& 3.672\,30$-$09& 4.225\,13$-$08\\
  & 5.091\,36$-$10& 1.561\,06$-$10& 7.539\,64$-$10& 3.276\,45$-$11& 3.681\,10$-$10& 8.026\,58$-$10\\
  & {\it 3.956\,26$-$11}& 1.688\,35$-$12& {\it 8.966\,33$-$11}& 1.656\,45$-$12& 2.884\,14$-$12& {\it 1.544\,45$-$10}\\
60& 1.832\,55$-$08& 1.376\,72$-$09& 3.313\,53$-$08& 9.415\,7$-$11& 4.285\,26$-$09& 4.371\,62$-$08\\
  & 4.477\,53$-$10& 1.445\,49$-$10& 6.503\,81$-$10& 3.195\,86$-$11& 3.352\,19$-$10& 6.741\,05$-$10\\
  & {\it 5.172\,91$-$11}& 1.193\,95$-$12& {\it 1.176\,66$-$10}& 1.360\,00$-$12& 1.884\,37$-$12& {\it 2.038\,55$-$10}\\
61& 1.933\,53$-$08& 1.617\,65$-$09& 3.456\,35$-$08& 1.328\,44$-$10& 4.953\,30$-$09& 4.490\,23$-$08\\
  & 3.902\,59$-$10& 1.330\,65$-$10& 5.541\,36$-$10& 3.109\,51$-$11& 3.025\,06$-$10& 5.559\,48$-$10\\
  & {\it 6.722\,11$-$11}& 8.034\,38$-$13& {\it 1.537\,73$-$10}& 1.094\,59$-$12& 1.138\,78$-$12& {\it 2.687\,22$-$10}\\
62& 2.030\,78$-$08& 1.884\,46$-$09& 3.585\,00$-$08& 1.818\,38$-$10& 5.679\,09$-$09& 4.574\,56$-$08\\
  & 3.366\,27$-$10& 1.216\,48$-$10& 4.651\,79$-$10& 3.015\,43$-$11& 2.699\,22$-$10& 4.481\,18$-$10\\
  & {\it 8.702\,42$-$11}& 5.061\,86$-$13& {\it 2.007\,02$-$10}& 8.590\,24$-$13& 6.155\,96$-$13& {\it 3.551\,46$-$10}\\
63& 2.123\,60$-$08& 2.180\,30$-$09& 3.696\,49$-$08& 2.436\,91$-$10& 6.464\,00$-$09& 4.615\,48$-$08\\
  & 2.868\,29$-$10& 1.102\,88$-$10& 3.834\,54$-$10& 2.910\,50$-$11& 2.373\,69$-$10& 3.505\,35$-$10\\
  & {\it 1.125\,29$-$10}& 2.911\,00$-$13& {\it 2.625\,03$-$10}& 6.521\,44$-$13& 2.804\,86$-$13& {\it 4.730\,03$-$10}\\
64& 2.211\,19$-$08& 2.509\,15$-$09& 3.786\,75$-$08& 3.218\,0$-$10& 7.304\,37$-$09& 4.598\,57$-$08\\
  & 2.408\,31$-$10& 9.895\,93$-$11& 3.088\,98$-$10& 2.789\,31$-$11& 2.046\,49$-$10& 2.630\,97$-$10\\
  & {\it 1.457\,69$-$10}& 1.464\,45$-$13& {\it 3.455\,25$-$10}& 4.728\,57$-$13& 9.553\,73$-$14& {\it 6.397\,43$-$10}\\
65& 2.292\,60$-$08& 2.875\,75$-$09& 3.849\,73$-$08& 4.203\,02$-$10& 8.178\,2$-$09& 4.497\,26$-$08\\
  & 1.985\,97$-$10& 8.762\,06$-$11& 2.414\,37$-$10& 2.640\,78$-$11& 1.713\,26$-$10& 1.856\,32$-$10\\
  & {\it 1.898\,57$-$10}& 5.956\,39$-$14& {\it 4.604\,33$-$10}& 3.200\,84$-$13& 1.803\,21$-$14& \\
66& 2.366\,50$-$08& 3.284\,65$-$09& 3.875\,22$-$08&  & & \\
  & 1.600\,87$-$10& 7.618\,61$-$11& 1.809\,79$-$10& & & \\
  & {\it 2.498\,29$-$10}& 1.652\,49$-$14& {\it 6.269\,60$-$10}& & & \\
67& 2.430\,79$-$08& & & & &\\
  & 1.252\,58$-$10&  &  &  &  &  \\
\end{longtable}
\end{center}

\begin{figure}
\begin{center}
\includegraphics[width=12cm]{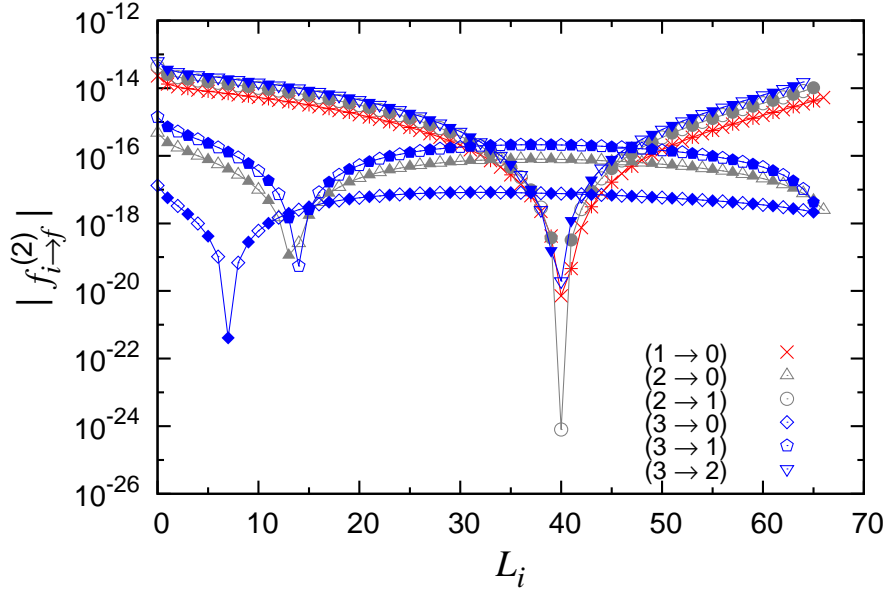}
\caption{Oscillator strengths for $L_f = L_i+2$ transitions.}
\label{fig:2}
\end{center}
\end{figure}

\begin{figure}
\begin{center}
\includegraphics[width=12cm]{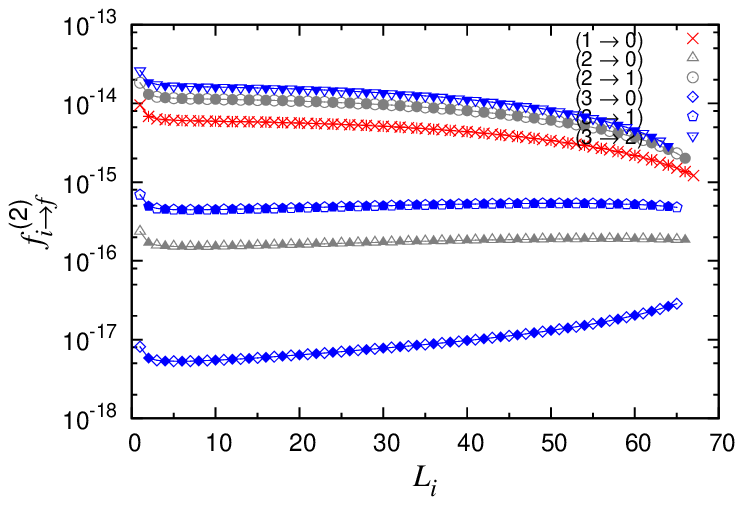}
\caption{Oscillator strengths for $L_f = L_i$ transitions.}
\label{fig:3}
\end{center}
\end{figure}

\begin{figure}
\begin{center}
\includegraphics[width=12cm]{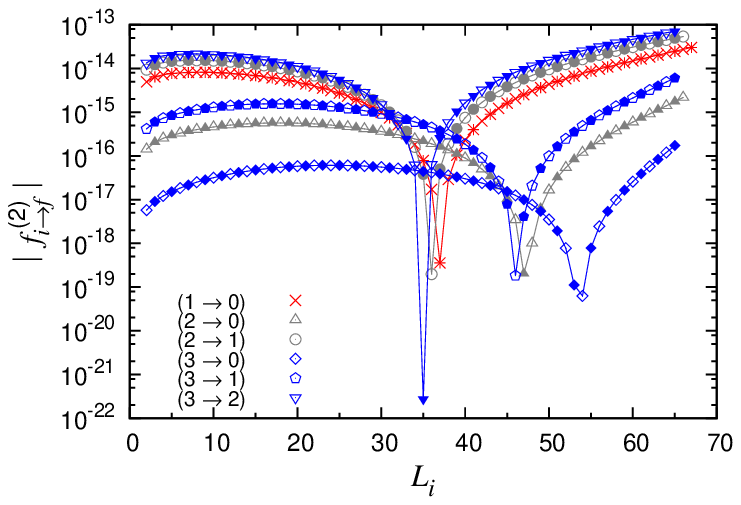}
\caption{Oscillator strengths for $L_f = L_i-2$ transitions.}
\label{fig:4}
\end{center}
\end{figure}

Quadrupole oscillator strength  are presented in Figure~\ref{fig:2}, \ref{fig:3} and \ref{fig:4}.  All the points connected by a curve  corresponds to transitions with the same value of the initial $v_i$ and final $v_f$ vibrational quantum number. For all the cases there is a systematic grouping of the curves depending of the difference between of the vibrational quantum numbers. Roughly the oscillator strength with $v_i-v_f=1$ are larger by an order of magnitude than those with  $v_i-v_f=2$ and $v_i-v_f=2$ are larger by an order of magnitude than those with  $v_i-v_f=3$.

Lifetimes for all states considered are calculated with 
\begin{equation}
\tau = \left(\sum_{E_f<E_i} W_{i  \rightarrow f}^{(2)}\right)^{-1},
\end{equation} 
and are displayed in Figure~\ref{fig:4}. Using the allowed electric quadrupole transitions presented in tables \ref{tab:2}  and \ref{tab:3} we obtain, for the vibrational states $(L=0)$  $v=1,2$ and $3$: $ \tau= 4.250\,039\times 10^7$\,s, $2.269\,525\times 10^7$\,s and $1.615\,861\times 10^7$\,s,   respectively.  The zero-vibrational states $(L^{\pi},0)$ have  the longer lifetime. Decreasing monotonically from $(2^{+},0)$  ($\tau= 777\,201$ years)  up to $L\sim 25$. In this domain $L\in[0-25]$, the lifetime for the excited  vibrational states  $v=1,2,3$  is around 9 months.  For $L>25$, the lifetime for all states decrease reaching a minimum around $L\sim 54$ ($\tau \sim 55$ days) and then starting to increase again slowly. 

\begin{figure}
\begin{center}
\includegraphics[width=12cm]{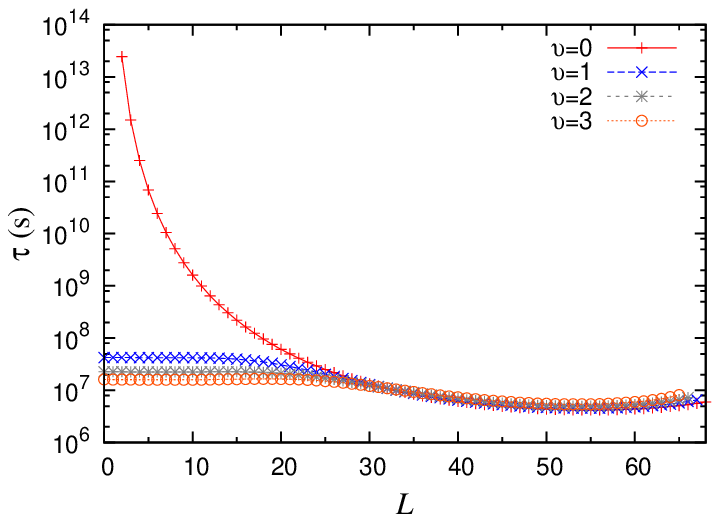}
\caption{Lifetimes  in seconds for the first four rotational bands $(v = 0-3)$.}
\label{fig:5}
\end{center}
\end{figure}

\section{Mass Dependence}

The mass dependence in the binding energy has been widely studied for three body Coulomb systems $m_1$, $m_2$ and $m_3$  and in particular for the symmetric configuration where $m_1=m_2$ (see for example \cite{GR1982} and \cite{RF2002} and references therein). The Lagrange-mesh method, briefly presented in section II, can easily be adapted to study the ground state energy of the symmetric three body systems with   $m_1=m_2=m$ and $m_3$. We consider the symmetric systems with one/two electrons: $^{\infty}$H$_2^+$ (infinite protons mass), Ps$^{-}$($e^+e^-e^-$), H$^-$, D$^-$, T$^-$, $^{\infty}$H$^-$ (infinite proton mass) and the muonic systems $\mu^+\mu^+e$ and $\mu^+e\,e$. In the Lagrange-mesh approach, energies and several properties of the systems Ps$^{-}$ and $^{\infty}$H$^-$  were  calculated in~\cite{HB1999}.  Results are presented in Table~\ref{tab:fit} together with those of the molecular ion H$_2^+$ and its isotopomers  D$_2^+$ and T$_2^+$. The values, in atomic units, of the particle masses  for the proton $m_p $, deuteron $m_d$, triton $m_t$ and muon $m_{\mu}$ are
\begin{eqnarray*}
m_p = 1836.152\,701,&\hspace{0.5cm} &m_d=3670.483\,014,\\
m_t = 5496.921\,58\,\,\,,& &m_{\mu}=206.768\,262\,.
\end{eqnarray*}

\begin{center}
\begin{table}[!thb]
\caption{Ground state energies for symmetric three body systems $m_1=m_2=m$ and $m_3$ with Coulomb interaction. $\beta=m_3/(2m+m_3)$. Results for H$_2^+$ and  D$_2^+$ from $^{a}$\cite{OPB12} and $^{b}$\cite{OP13}. The symbol "$\infty$" stands for an infinitely heavy particle. Comparison is made with $^{c}$\cite{YZL03}, $^{d}$\cite{HBGD2000}, $^{e}$\cite{HNN2009}, $^{f}$\cite{SBGS1998},  $^{g}$\cite{Fro2005}, $^{h}$\cite{FR2004}, $^{i}$\cite{FS2003} ,  $^{j}$\cite{Nak2007} (rounded).  }
\label{tab:fit}
\begin{tabular}{cllll}
\hline\hline
System   & $m$&  $\beta$             & $E_t$\,(present)  &  Reference\\
\hline
$^{\infty}$H$_2^+$ & $\infty$&0.0                &-0.602\,634\,619\,105 & \\  
\,\,\,\,\,\,T$_2^+$& 5496.921\,58&9.0951729$\times10^{-5}$&-0.599\,506\,910\,111\,5 &-0.599\,506\,910\,111\,541$^{c}$\\
\,\,\,\,\,\,D$_2^+$& 3670.483\,014&1.3620330$\times10^{-4}$&-0.598\,788\,784\,330\,7$^{a}$ &-0.598\,788\,784\,330\,68$^{d}$ \\
\,\,\,\,\,\,H$_2^+$& 1836.152\,701&2.7223437$\times10^{-4}$&-0.597\,139\,063\,123\,41$^{b}$  &-0.597\,139\,063\,123\,405$^{e}$\\
\,\,\,\,$\mu^+\mu^+e$& 206.768\,262&2.4123327$\times10^{-3}$&-0.585\,126\,097\,219\,20&-0.585\,126\,097\,219\,193$^{f}$\\
\,\,\,\,Ps$^{-}$         & 1.0 &0.33333333&-0.262\,005\,070\,232\,97 & -0.262\,005\,070\,232\,980\,$^{g}$\\
$\mu^+e\,e$           & 1.0&0.99042000&-0.525\,054\,806\,243\,53&-0.525\,054\,806\,243\,526\,$^{h}$\\
\,\,\,\,\,\,H$^-$         & 1.0&0.99891195 &-0.527\,445\,881\,114\,18 &-0.527\,445\,881\,114\,179\,$^{i}$\\
\,\,\,\,\,\,D$^-$         & 1.0&0.99945541&-0.527\,598\,324\,686\,48&  -0.527\,598\,324\,686\,478\,$^{i}$\\ 
\,\,\,\,\,\,T$^-$         & 1.0&0.99963629&-0.527\,649\,048\,203\,01&  -0.527\,649\,048\,202\,999\,95\,$^{i}$\\ 
\,\,$^{\infty}$H$^-$ & 1.0&1.0 &-0.527\,751\,016\,544\,38 &             -0.527\,751\,016\,544\,377\,$^{i,j}$\\
\hline\hline
\end{tabular}
\end{table}
\end{center}

Following Gur'yanov and Rebane~\cite{GR1982} the ground state energy is expanded as follows
\begin{equation}
\label{expn}
E(\beta)  \approx m\beta f(\beta)\,,
\end{equation}
where
\begin{equation}
\label{expnf}
f(\beta)= \sum_{j=0}^{n} C_j \beta^{j/2}\,,
\end{equation}
and  $\beta=m_3/(2m+m_3)$ varies between $[0,1]$. The parameter $C_0$ is fixed in terms of  the value of the ground state energy of the static hydrogen molecular ion $^{\infty}$H$_2^+$. First  $5$ terms in (\ref{expn}), $n=5$ reproduce no less than 7 s.d. in energies with parameters 
\begin{center}
\begin{tabular}{lll}
$C_1 =  0.64179582$,&$C_2 =  0.28372246$,&$C_3 = -0.15304332$,\\
$C_4 = -0.21968058$,&$C_5 =  0.12472387$,&
\end{tabular}
\end{center}
which are in agreement to the parameters found in~\cite{GR1982}. Increasing number of terms up to $7$, we reproduce no less than $10$ s.d. in energy (see Table~\ref{tab:fit}) with parameters 
\begin{center}
\begin{tabular}{lll}
$C_1 =  0.64177988217$,& $C_2 =  0.28534422820$,& $C_3 = -0.18731634808$,\\
$C_4 = -0.04273755179$,& $C_5 = -0.24388850969$,& $C_6=0.33561102864$,\\
$C_7=-0.11127450779$. &&\\
\end{tabular}
\end{center}

The expansion $f(\beta)$ (\ref{expnf})  is an increasing function of $\beta \in [0,1]$ and is presented in Figure~\ref{fig:5}.
\begin{figure}
\begin{center}
\includegraphics[width=12cm]{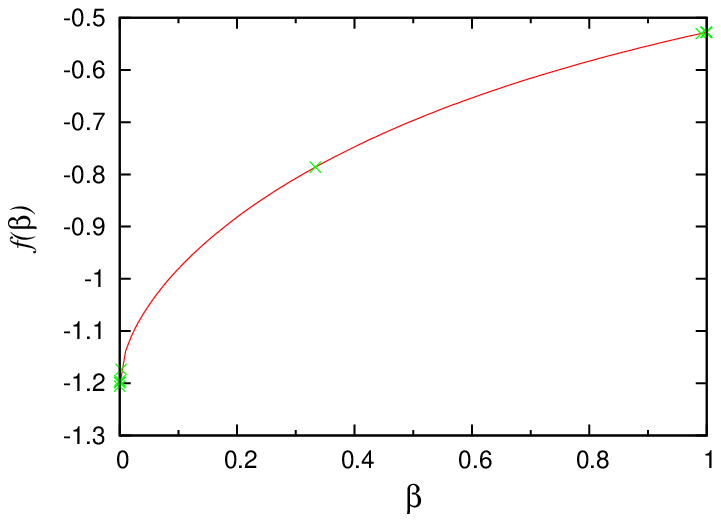}
\caption{Expansion $f(\beta)$ (\ref{expnf}) for $n=7$ (line). Symbols indicate  all  $11$ systems present in Table~\ref{tab:fit}.}
\label{fig:5}
\end{center}
\end{figure}

\section*{Conclusions}
\label{s4}
Summarizing, in order to solve the non-relativistic Shr$\ddot{\rm o}$dinger equation for the three body system with Coulomb interaction composed of two tritons and one electron, the  Lagrange-mesh method  is applied.  The spectra for the four lowest vibrational sates $v=0,1,2,3$  is presented with $13$, $12$, $11$ and $10$ significant digits using $54$ mesh points in the $x-y$ perimetric coordinates and $18$ mesh points in the $z$ perimetric coordinate.  It is found that the vibrational states $v=0,1,2,3$ support $62$, $61$, $60$ and $58$ rotational states, respectively. Some quasi bound states are also given.  The vibrationless band $(L^{\pi},0)$ supports $27$ ($11$) rotational bound states more than the  isotopomer H$_2^+$~\cite{OPB12}  (D$_2^+$ ~\cite{OP13}). Table~\ref{tab:4} presents a complete comparison between these systems.  

Using the wave functions provided by the Lagrange mesh method it is easy to calculate the electric quadrupole transitions probabilities per second $W^{(2)}_{i\rightarrow f}$. All possible transitions probabilities are presented with six significant digits. Quadrupole oscillator strength $f^{(2)}_{i\rightarrow f}$ for $\Delta L = 2, 0,-2$ are depicted in Figures \ref{fig:2}, \ref{fig:3} and \ref{fig:4}, respectively. For $L_i = 41$  and beyond the $(L_i, 1 )\rightarrow (L_i + 2, 0)$ transitions are replaced by $(L_i + 2, 0) \rightarrow (L_i, 1 )$ transitions. The rotational state $L_i$ after which the initial and final states are exchanged for the three cases $v_{1\rightarrow 0}$, $v_{2\rightarrow 1}$ and $v_{3\rightarrow 2}$ and comparison with the molecular ions H$_2^+$ and D$_2^+$ are shown  in Table~\ref{tab:4}.
 
\begin{center}
\begin{table}[!thb]
\caption{Comparison between the isotopomers H$_2^+$, D$_2^+$ and T$_2^+$. $a)$ Number of rotational bound states for the  lowest three vibrational states and  $b)$ rotational state after which the initial and final states are exchanged.}
\label{tab:4}
\begin{tabular}{lllcccccccc}
\hline
  &         & &\multicolumn{4}{c}{Rotational bound states}&&\multicolumn{3}{c}{Change direction}\\  \cline{4-7} \cline{9-11}  
  &Mass &$(0^+,0)$&$v=0$&$v=1$&$v=2$&$v=3$&&$v_{1\rightarrow 0}$&$v_{2\rightarrow 1}$&$v_{3\rightarrow 2}$\\
\hline
H$_2^+$&$1836.152\,701$&$-0.597\,139\,063\,123\,3$&35&34&33&31&& 23 & 23 & 22\\
D$_2^+$&$3670.483\,014$&$-0.598\,788\,784\,330\,7$&51&49&48&47&& 33 & 33 & 32\\
T$_2^+$&$5496.921\,58$  &$-0.599\,506\,910\,111\,5$&62&61&60&58 && 41 & 41 & 40\\
\hline
\end{tabular}
\end{table}
\end{center}
 
Qualitatively the behaviour of the electric quadrupole transitions $W^{(2)}$ resembles to that of the electric quadrupole oscillator strength (see Figures \ref{fig:2}, \ref{fig:3} and \ref{fig:4}).  Comparing with the isotopomers H$_2^+$~\cite{OPB12} and D$_2^+$ ~\cite{OP13}  the hierarchy  
\begin{equation}
W_{{\rm H}_2^+}^{(2)} > W_{{\rm D}_2^+}^{(2)} > W_{{\rm T}_2^+}^{(2)},
\end{equation}
is found, except at some points  near to the minima  and at the ends of the curves. As a consequence, the lifetimes of T$_2^+$ are larger than those of D$_2^+$ and H$_2^+$,  around three times and 21 times, respectively.

In order to study the mass dependence for  the ground state of three body systems, we consider the systems $^{\infty}$H$_2^+$, $^{\infty}$H$^-$,  Ps$^{-}$, H$^-$, D$^-$, T$^-$, $\mu^+\mu^+e$, $\mu^+e\,e$  and those previously investigated H$_2^+$ and  D$_2^+$ (see Table \ref{expn}).  Expanding the energy as (\ref{expn}) and keeping $7$ terms, $n=7$, we reproduce no less than $10$ s.d.

\begin{acknowledgements}
I would like to thank Professor A Turbiner for his valuable comments and suggestions, to  CONACyT (Mexico) for a postdoctoral grant and  the Instituto de Ciencias Nucleares (UNAM, Mexico City) for their kind hospitality.  This work was partially supported by the University Program FENOMEC (UNAM, Mexico).
\end{acknowledgements}


\end{document}